\documentclass[a4paper,11pt]{article}
\pdfoutput=1 

\usepackage[utf8]{inputenc}

\usepackage{jcappub} 

\usepackage[T1]{fontenc} 
\usepackage[utf8]{inputenc}
\usepackage{siunitx}
\usepackage{url}
\usepackage{tensor}
\usepackage{float}
\usepackage{mathtools}
\usepackage{amssymb}
\usepackage{amsmath}
\usepackage{makecell,tabularx,multirow}
\usepackage{color, colortbl}
\usepackage{adjustbox}
\usepackage[pages=some]{background}
\usepackage{hyperref}
\usepackage{accents}
\usepackage{makecell}
\usepackage[margin=1in]{geometry}
\usepackage{stackengine}

\usepackage[font=footnotesize]{caption}

\hypersetup{
    colorlinks=true,
    linkcolor=blue,
    filecolor=magenta,
    citecolor=red
}
\usepackage{comment}
\usepackage[normalem]{ulem}

\usepackage[caption=false]{subfig}

\newcolumntype{C}{>{\centering\arraybackslash}X}
\newcolumntype{x}[1]{>{\centering\arraybackslash\hspace{0pt}}p{#1}}

\newcommand{\udt}[3]{#1^{#2}_{\phantom{#2}#3}}

\begin{document}

\title{\textbf{Spatial Dependence of the Growth Factor in Scalar-Tensor Cosmology}}

\author[a,b,1]{Maria Caruana,\note{Corresponding author}}
\author[a,b]{Gabriel Farrugia}
\author[a,b]{Jackson Levi Said,}
\author[c]{Joseph Sultana}

\affiliation[a]{Institute of Space Sciences and Astronomy, University of Malta, Msida, Malta.}
\affiliation[b]{Department of Physics, University of Malta, Msida, Malta.}
\affiliation[c]{Department of Mathematics, University of Malta, Msida, Malta.}

\emailAdd{maria.caruana.16@um.edu.mt}
\emailAdd{gfarr02@um.edu.mt}
\emailAdd{jackson.said@um.edu.mt}
\emailAdd{joseph.sultana@um.edu.mt}

\abstract{
Scalar-tensor theories have taken on a key role in attempts to confront the growing open questions in standard cosmology. It is important to understand entirely their dynamics at perturbative level including any possible spatial dependence in their growth of large scale structures. In this work, we investigate the spatial dependence of the growth rate of scalar-tensor theories through the M\'{e}sz\'{a}ros equation. We confirm that at subhorizon level this dependence does not play a major role for viable models. However, we establish conditions on which this criterion is met which may be important for developing new models. In our work, we consider three specific models that exhibit spatial dependence of the growth rate at subhorizon modes, which may also be important for early Universe models.}

\maketitle

\section{Introduction}\label{sec:intro}
The standard model of cosmology rests on a wealth of decades of observational evidence both at cosmic and astrophysical scales \cite{Misner:1974qy,Clifton:2011jh} wherein cold dark matter (CDM) operates to stabilize compact gravitational systems \cite{Baudis:2016qwx,Bertone:2004pz}, while dark energy materializes as a cosmological constant \cite{Peebles:2002gy,Copeland:2006wr}. Despite these successes, internal consistency issues have persisted in the cosmological constant $\Lambda$ description of dark energy \cite{Weinberg:1988cp}, and the possibility of directly measuring dark matter particles remains out of reach \cite{Gaitskell:2004gd}. In the last few years another issue has arisen where the standard model of cosmology, or $\Lambda$CDM, has been called into question. The problem has to do with the present value of the expansion rate of the Universe, the Hubble constant $H_0$, where a statistically significant tension has appeared between the value of the Hubble constant as fit using some different surveys \cite{DiValentino:2020zio}. One viewpoint of the discrepancy is that the tension is between cosmology-independent late time, or local, measurements of the Hubble constant \cite{Riess:2019cxk,Shajib:2023uig}, and those based on predictions of the Hubble constant using early time observations \cite{Aghanim:2018eyx,Ade:2015xua}. While other measurements may have an important contribution to defining the contours of the broader cosmological tensions problem \cite{Baker:2019nia,2017arXiv170200786A,Barack:2018yly}, the existence of this issue may point to the possibility of new physics \cite{Abdalla:2022yfr}.

The reaction to the growing problem of cosmological tensions in $\Lambda$CDM has taken several forms with the possibility of it being due to some underlying systematic that has influenced the plethora of surveys seems progressively unlikely. As for the underlying physics, there have been several promising proposals for beyond $\Lambda$CDM cosmology including novel physics contained before recombination \cite{Poulin:2023lkg}, extra degrees of freedom in the neutrino sector \cite{DiValentino:2021imh}, and global modifications to the governing gravitational physics \cite{Addazi:2021xuf,CANTATA:2021ktz,Bahamonde:2021gfp,Bamba:2012cp, Nojiri:2010wj,Nojiri:2017ncd}. Scalar-tensor theories \cite{Joyce:2016vqv,Palti:2019pca,Capozziello:2007ec} have featured in all these possible avenues for confronting the cosmic tensions problem. They are the simplest form of modification one can apply for any base model such as $\Lambda$CDM. These continue to be the principle method by which cosmic inflation is described \cite{Guth:1980zm,Linde:1981mu} where a scalar field is dynamical in the extremely early Universe and used to resolve several cosmological problems including the horizon and flatness problems. 

Scalar-tensor theories can also be explored collectively through Horndeski gravity \cite{Horndeski:1974wa,Horndeski:2024sjk,Kobayashi:2019hrl} wherein a single scalar field is considered through its most general second-order realization. This formulation of scalar-tensor theories, while complex, gives a concise expression through which to probe different phenomenology. However, recent constraints on the speed of propagation of gravitational waves has put extreme limits on the more exotic parts of the theory \cite{LIGOScientific:2017zic,Ezquiaga:2017ekz}. While one can consider other avenues to revive Horndeski gravity including non-Riemannian geometries \cite{Bahamonde:2019shr,Bahamonde:2022cmz} and beyond Horndeski theories \cite{Gleyzes:2014dya,Traykova:2019oyx}, this has placed severe constraints on the model space of the original theory. Within this reality of viability model space, we consider a broadly general minimally coupled scalar field class of models on which to examine the growth of structure and its relation to the growth index. This is important since the cosmic tensions problem has also been connected to the evolution of the growth of large scale structures \cite{DiValentino:2020vvd,Poulin:2022sgp,Kazantzidis:2018rnb}. As more survey results are made available, it will become all the more important to have a deeper understanding of the evolution of growth of structures as well as their $k$-dependence.

Our objective in this work is to probe the growth of large scale structure in promising models of scalar-tensor theories from the literature \cite{Felder:2002jk,Linde:1983gd,Copeland:1997et}. We do this by calculating the M\'esz\'aros equation for this class of models. This is performed without setting the subhorizon limit initially so that we can probe the $k$-dependence of the evolving modes. Moreover, we consider the power-law growth rate dependence on the matter density parameter approximation \cite{Linder:2005in}. The result is a system that can be solved to express the evolution of different modes in the growth of large scale structures. In this work, we consider three representative models that appear in the literature. Through these models we analyze the evolution of the equation of state for each model as well as the fractional difference of the growth factor against the baseline $\Lambda$CDM model. We use these solutions to explore the $k$-dependence and subhorizon limits for each scalar-tensor model. We start by briefly describing the underlying background and perturbative cosmology in Sec.~\ref{sec:background}. This leads naturally to the M\'esz\'aros equation and the expression of the growth index equations in Sec.~\ref{sec:meszaros}. We then implement the models we consider and solve them numerically in Sec.~\ref{sec:models}. Finally, we close with a summary of our main results and a discussion of possible future work in Sec.~\ref{sec:conclu}.

\section{Background and Perturbation Cosmology} \label{sec:background}

Consider the modified action where, in addition to the Einstein-Hilbert action, there is a potential and kinetic terms given by
\begin{align} \label{eq:action}
    \mathcal{S} \equiv \frac{1}{2\kappa^2} \int \text{d}^4x \sqrt{-g} \, (R + V(\phi) + P(\phi)X) + \int \text{d}^4x \sqrt{-g} \, \mathcal{L}_{m}\,,
\end{align}
where $g$ is the metric determinant, $V$ and $P$ are arbitrary functions of the scalar field $\phi$, kinetic term $X = -\frac{1}{2} \partial_{\mu}\phi\partial^{\mu}\phi$, $\kappa^2 = 8 \pi G$ and $\mathcal{L}_{m}$ is the matter Lagrangian. At background, the spatially flat Friedmann-Lema\^itre-Robertson-Walker (FLRW) metric in terms of Cartesian coordinates~\cite{Krssak:2018ywd}
\begin{align} \label{eq:FLRW_metric}
    \text{d}s^2 = \text{d}t^2 - a(t)^2(\text{d}x^2 + \text{d}y^2 + \text{d}z^2)\,,
\end{align}
where $a(t)$ is the scale factor is applied. Variation with respected to the metric field yields the field equations
\begin{align} \label{eq:tetrad_variation}
    W_{\alpha\beta}: &\qquad \kappa^2 \theta_{\alpha\beta} = G_{\alpha\beta} - \frac{1}{2} g_{\alpha\beta} \left(V(\phi) - \frac{1}{2} P(\phi) \partial_{\mu}\phi\, \partial^{\mu}\phi\right) - \frac{1}{2} P(\phi) \partial_{\alpha}\phi\, \partial_{\beta}\phi\,,
\end{align}
where $G_{\alpha\beta}$ is the Einstein tensor, $\phi$ subscript represent partial derivatives with respect to the scalar field and $\theta_{\alpha\beta}$ is the matter energy-momentum tensor~\cite{Saez-Gomez:2016wxb}
\begin{align}\label{eq:energy_momentum_equation}
    \theta^{\alpha\beta} = (\rho + p) u^{\alpha}u^{\beta} - p g^{\alpha\beta} - \Pi^{\alpha\beta}\,,    
\end{align}
where $u_{\alpha}$ is the fluid four-velocity vector and $\Pi_{\alpha\beta}$ is the anisotropic stress tensor accounting for directional dependency of pressure. Additionally, taking variation with respect to the scalar field gives
\begin{align}\label{eq:field_variation}
    W_{\phi}: &\qquad 0 = V_{\phi} + \frac{1}{2} P_{\phi} \partial_{\mu}\phi\, \partial^{\mu}\phi + P \partial_{\mu}\partial^{\mu}\phi + P \partial_{\mu}(\sqrt{-g}) \partial^{\mu}\phi\,.
\end{align}
Next, we will consider the following background and perturbation variables to construct our system of equations. First and foremost, the scalar field is perturbed around an only time dependent background given by
\begin{align}\label{eq:field_perturbation}
    \phi \rightarrow \phi(t) + \delta\phi(t,x,y,z)\,,
\end{align}
where the perturbation is taken to be temporal and spatial. The metric is given by
\begin{align}\label{eq:metric_perturbation}
    g_{\alpha\beta} \rightarrow 
    \begin{bmatrix}
    1  & 0\\
    0 & -a^2\delta_{ij}
    \end{bmatrix}
    +
    \begin{bmatrix}
    2 \varphi & a \partial_{i}B\\
    a \partial_{i}B & 2a^2(\psi\delta_{ij}+ \partial_{i}\partial_{j}E)\,,
    \end{bmatrix}\,, 
\end{align}
where $\{\varphi, B,\psi, E\}$ are gravitational scalar perturbations to form a generalized perturbed metric. In particular, we can consider the longitudinal gauge by setting $B = 0$ and $E = 0$. The energy-momentum tensor background and linear order is given by~\cite{Zheng:2010am,Farrugia:2016pjh}
\begin{align}\label{eq:energy_momentum_perturbation}
    \theta_{\alpha\beta} = \begin{bmatrix}
        \rho & 0\\
        0 & a^2 p\,\delta_{ij}
    \end{bmatrix}
    +
    \begin{bmatrix}
        \delta\rho & -a^2(\rho + p) \partial_{i}v\\
        -a^2(\rho + p) \partial_{i}v & a^2 (\delta p \,\delta_{ij} + \partial_{i}\partial_{j}\pi^{\text{s})}
    \end{bmatrix}\,,
\end{align}
obtained by extending Eq.~\eqref{eq:energy_momentum_equation} where $v$ is the scalar component of the velocity vector $v_{i} = \tfrac{u_i}{u_0}$ and $\pi^{\text{s}}$ is the scalar component of the anisotropic stress satisfying the properties $\udt{\Pi}{0}{0} = \udt{\Pi}{0}{i} = u^{\mu}\Pi_{\mu\nu} = 0$. Applying the above extensions, we can obtain the background equations from Eqs~(\ref{eq:tetrad_variation}-\ref{eq:field_variation}) 
\begin{align}
    \label{eq:field_equations_background_W00} 
    W_{00}: \qquad\,\,\,\,\, 2 \kappa^{2} \rho &= 6H^2 - V + P X \,, \\
    \label{eq:field_equations_background_Wii} 
    W_{ii}: \qquad -2\kappa^{2}p &= 6 H^2 + 4 \dot{H} - V - PX \,, \\
    \label{eq:field_equations_background_Wphi} 
    W_{\phi}: \qquad\qquad\,\, 0 &= V_{\phi} - X P_{\phi} + 3 H P \dot{\phi} + P \ddot{\phi}\,, 
\end{align}
where $H = \tfrac{\dot{a}}{a}$ is the Hubble parameter. Comparing the Friedmann equations~(\ref{eq:field_equations_background_W00}-\ref{eq:field_equations_background_Wii}) to their GR counterparts yields the effective dark energy portion of the field equations
\begin{align}
    \label{eq:eff_ener_den} 2 \kappa^2 \rho_{\text{DE}} &= - V + P X\,,\\
    \label{eq:eff_pressure} -2 \kappa^{2} p_{\text{DE}} &= 2 \kappa^2 \rho_{\text{DE}} -2 P X \,,
\end{align}
such that the effective equation of state is given by
\begin{align}\label{eq:effective_EoS}
    \omega_{\text{DE}} = \frac{p_{\text{DE}}}{\rho_{\text{DE}}} = - 1  - \frac{ 2 P X }{ V - P X} \,,
\end{align}
where the $\Lambda$CDM limit is obtained when eliminating $P$ or $X$ contributions to obtain $\omega_{\text{DE}}=-1$. From this point forward all expressions are written in terms of spatial Fourier transformation. Additionally, the linear field equations are
\begin{align}
    \label{eq:field_equations_perturbations_W00} \delta W_{00}: \quad 0 &= \kappa^2 \delta\rho + (-\tfrac{1}{2} P \dot{\phi}^2 + 6 H^2) \varphi+ 2\tfrac{k^2}{a^2}\psi + \tfrac{1}{4}(P_{\phi} \dot{\phi}^2 + 2 V_{\phi}) \delta\phi + 6 H \dot{\psi}
   + \tfrac{1}{2} P \dot{\phi} \,\dot{\delta\phi}  \,,\\
    \label{eq:field_equations_perturbations_W0i} \delta W_{(0i)}: \quad 0 &= 
    \kappa^{2}a^2 (\rho + p) v + 2 H \varphi - \tfrac{1}{2} P \dot{\phi} \delta\phi + 2 \dot{\psi} \,, \\
    \label{eq:field_equations_perturbations_Wij} \delta W_{ij}: \quad 0 &= \varphi - \psi \,, \\
    \label{eq:field_equations_perturbations_Wii} \delta W_{ii}: \quad 0 &= - \kappa^{2} \delta p - \tfrac{2}{3} \tfrac{k^2}{a^2} (\varphi -\psi) + ( 6 H^{2} + \tfrac{1}{2} P \dot{\phi}^{2} + 4 \dot{H}) \varphi + \tfrac{1}{4}( 2 V_{\phi} - P_{\phi}\dot{\phi}^2 ) \delta\phi \nonumber \\ &\quad  - \tfrac{1}{2} P \dot{\phi} \dot{\delta\phi} + 6 H \dot{\psi} + 2 (H \dot{\varphi} + \ddot{\psi}) \\
    \label{eq:field_equations_perturbations_Wphi} \delta W_{\phi}: \quad 0&= 2 V_{\phi}\varphi + (\tfrac{1}{2} P_{\phi\phi} \dot{\phi}^2 + V_{\phi\phi} - \tfrac{P_{\phi}}{2P}(P_{\phi} \dot{\phi}^2 + 2 V_{\phi}) +\tfrac{k^2}{a^2} P) \delta\phi - P \dot{\phi} \,\dot{\varphi}  -3 P \dot{\phi} \dot{\psi} \nonumber \\ &\quad + P\,\ddot{\delta\phi} + (3 H P + P_{\phi}\dot{\phi})\dot{\delta\phi}   \,.
\end{align}
It should be noted that $W_{i0}=W_{0i}$ as required by the symmetry of the field equations. Moreover, the conservation of the energy momentum tensor~\cite{misner1973gravitation}
\begin{align}\label{eq:conservation_energy_momentum}
    \nabla_{\mu}\udt{\theta}{\mu}{\nu} =\partial_{\mu}\udt{\theta}{\mu}{\nu} + \Gamma^{\mu}_{\mu\alpha} \udt{\theta}{\alpha}{\nu} - \Gamma^{\alpha}_{\mu\nu} \udt{\theta}{\mu}{\alpha} = 0\,,
\end{align}
gives the background continuity equation
\begin{align} \label{eq:continuity_backgorund}
    \dot{\rho}+3H(\rho+p) = 0\,,
\end{align}
which can be also be obtained by manipulating the Friedmann equations~(\ref{eq:field_equations_background_W00}-\ref{eq:field_equations_background_Wii}) and their time derivatives. The linearized version of Eq.~\eqref{eq:conservation_energy_momentum} yields the continuity and velocity (Euler) equations:
\begin{align}
    \label{eq:continuity_perturbation} 0 &= 3 H(\delta p + \delta\rho) + \dot{\delta\rho} - (p+\rho)(k^2v + 3\dot{\psi}) - H k^2 \pi^{\text{s}}\,,\\
    \label{eq:velocity_perturbation} 0 &= (p+\rho)\varphi + \delta p + a^{2}[(2 H (\rho + p)+\dot{p})v + (\rho + p)\dot{v}] - k^2 \pi^{\text{s}}\,.
\end{align}
The combination of first order gravitational Eqs~(\ref{eq:field_equations_perturbations_W00}-\ref{eq:field_equations_perturbations_Wphi}) and matter Eqs~(\ref{eq:continuity_perturbation}-\ref{eq:velocity_perturbation}) perturbation equations and their background equations~(\ref{eq:field_equations_background_W00}-\ref{eq:field_equations_background_Wphi}) provides the system of equations to explain the evolution of the scalar perturbations on which our framework is built.

\section{M\'esz\'aros Equation, Growth Factor and Growth Index} \label{sec:meszaros}

We are interested in looking into the impact of the scalar field $\phi$ on the growth of cold dark matter (CDM) content. It should be noted that in Refs~\cite{Sharma:2021fou,Sharma:2021ivo}, the scalar field is treated as a separate dark energy energy component interacting with pressureless CDM, while in this work all matter is taken to be CDM. For this reason we, can assume
\begin{align} \label{eq:CDM_conditions}
    p = \delta p = 0\,,
\end{align}
to consider the epoch deep within matter dominated era up to current time, and where the radiation epoch is ignored. Additionally, the fluid's anisotropic stress is omitted from Eq.~\eqref{eq:energy_momentum_perturbation} because quadrupole contributions are insignificant when considering CDM~\cite{Dodelson:2003ft}. We define matter perturbation in terms of the gauge invariant comoving fractional matter perturbation
\begin{align} \label{eq:fractional_matter_perturbation}
    \delta_{m} := \frac{\delta\rho}{\rho} - 3 a^2 H v\,.
\end{align}
To obtain the M\'esz\'aros equation, we require to set up a system of equations using field equations~(\ref{eq:field_equations_perturbations_W00}-\ref{eq:field_equations_perturbations_Wphi}) and conservation of energy equations~(\ref{eq:continuity_perturbation}-\ref{eq:velocity_perturbation}). We opt to work in terms of $k$-dependencies i.e. without applying the subhorizon limit $k \gg aH$ at the initial stages of the calculation. In $f(R)$ theories it has been shown that the M\'esz\'aros equation obtained through the subhorizon limit and ignoring time-dependencies~\cite{Tsujikawa:2007gd,DeFelice:2010aj-f(R)_theories} leads to a discrepancy from the result obtained in Ref.~\cite{delaCruz-Dombriz:2008ium}. Although this is not necessarily always the case, as shown in teleparallel $f(T)$ gravity~\cite{Capozziello:2023giq}, it is ideal to consider a more generalised approach. The initial form of M\'esz\'aros equation can be obtained from the continuity and velocity equations~(\ref{eq:continuity_perturbation}-\ref{eq:velocity_perturbation}), rewritten as
\begin{align}
    \label{eq:continuity_mod} 0 &= a^2 (- \tfrac{k^2}{a^2} + 3 \dot{H} + 6 H^2) v  + 3 a^2 H \dot{v} + \dot{\delta}_{m} - 3 \dot{\psi}\,, \\
    \label{eq:velocity_mod} 0 &= \varphi + a^2 (2  H v +  \dot{v})\,. 
\end{align}
where Eq.~\eqref{eq:fractional_matter_perturbation} have been applied. Hence, substituting Eq.~\eqref{eq:velocity_mod} into Eq.~\eqref{eq:continuity_mod} yields to an expression for $v$:
\begin{align} \label{eq:v_equation}
    a^2 (\tfrac{k^2}{a^2} - 3 \dot{H} ) v = - 3 H\varphi + \dot{\delta}_{m} - 3 \dot{\psi}\,. 
\end{align}
Thus, using Eq.~\eqref{eq:v_equation}, a system of equations can be constructed from the linearized field equations~(\ref{eq:field_equations_perturbations_W00}-\ref{eq:field_equations_perturbations_Wphi}) and their time derivatives to obtain functions of the form:
\begin{align}
    \label{eq:A1} \delta W_{00}: \qquad 0 &= \mathcal{A}_{1}(\delta\phi, \dot{\delta\phi}, \varphi, \psi, \dot{\psi}, \delta_{m}, \dot{\delta}_{m})\,, \\
    \label{eq:A2} \delta W_{0i}: \qquad 0 &= \mathcal{A}_{2}(\delta\phi, \varphi, \dot{\psi},  \dot{\delta}_{m})\,, \\
    \label{eq:A3} \delta W_{ii}: \qquad 0 &= \mathcal{A}_{3}(\delta\phi, \dot{\delta\phi}, \varphi, \dot{\varphi},  \psi, \dot{\psi}, \ddot{\psi})\,, \\
    \label{eq:A4} \delta W_{ij}: \qquad 0 &= \mathcal{A}_{4}(\varphi, \psi)\,, \\
    \label{eq:A5} \delta W_{\phi}: \qquad 0 &= \mathcal{A}_{5}(\delta\phi, \dot{\delta\phi}, \ddot{\delta\phi}, \varphi, \dot{\varphi}, \dot{\psi})\,, \\
    \label{eq:A6} \dot{\delta W}_{00}: \qquad 0 &= \mathcal{A}_{6}(\delta\phi, \dot{\delta\phi}, \ddot{\delta\phi},  \varphi, \dot{\varphi}, \psi, \dot{\psi}, \ddot{\psi}, \delta_{m}, \dot{\delta}_{m}, \ddot{\delta}_{m})\,, \\
    \label{eq:A7} \dot{\delta W}_{0i}: \qquad 0 &= \mathcal{A}_{7}(\delta\phi, \dot{\delta\phi}, \varphi, \dot{\varphi}, \dot{\psi}, \ddot{\psi}, \dot{\delta}_{m}, \ddot{\delta}_{m})\,, \\
     \label{eq:A8} \dot{\delta W}_{ij}: \qquad 0 &= \mathcal{A}_{8}(\varphi, \dot{\varphi}, \psi, \dot{\psi})\,,    
\end{align}
which yields 8 linearly independent equations corresponding to the 8 gravitational perturbed variables. Further details of the $\mathcal{A}_i$ functions is given in Appendix~\ref{sec:AppendixA}. Thus, each gravitational variable can be expressed in terms of $\delta_{m}$ and its time derivatives, i.e.
\begin{align} \label{eq:matrix}
    \overbracket{\begin{pmatrix}
        \mathcal{B}_{11} & \mathcal{B}_{12} & 0 & \mathcal{B}_{14} & 0 & \mathcal{B}_{16} & \mathcal{B}_{17} & 0 \\
        \mathcal{B}_{21} & 0 & 0 & \mathcal{B}_{24} & 0 & 0 & \mathcal{B}_{27} & 0\\
        \mathcal{B}_{31} & -\tfrac{1}{2} P \dot{\phi} & 0 & \mathcal{B}_{34} & 24 H & \mathcal{B}_{36} & 72 H & 24\\
        0 & 0 & 0 & 1 & 0 & -1 & 0 & 0\\
        \mathcal{B}_{51} & \mathcal{B}_{52} & P & 2V_{\phi} & -P\dot{\phi} & -3P\dot{\phi} & \mathcal{B}_{57} & \mathcal{B}_{58}\\
        \dot{\mathcal{B}}_{11} & \mathcal{B}_{11}+\dot{\mathcal{B}}_{12} & \mathcal{B}_{12} & \dot{\mathcal{B}}_{14} & \mathcal{B}_{14} & \dot{\mathcal{B}}_{16} & \mathcal{B}_{16} + \dot{\mathcal{B}}_{17}& \mathcal{B}_{17}\\
        \dot{\mathcal{B}}_{21} & \mathcal{B}_{21} & 0 & \dot{\mathcal{B}}_{24} & \mathcal{B}_{24} & 0 & \dot{\mathcal{B}}_{27} & \mathcal{B}_{27}\\
        0 & 0 & 0 & 0 & 1 & 0 & -1 & 0
    \end{pmatrix}}^{\mathcal{B}}
    \overbracket{\begin{pmatrix}
        \delta\phi\\
        \dot{\delta\phi}\\
        \ddot{\delta\phi}\\
        \varphi\\
        \dot{\varphi}\\
        \psi\\
        \dot{\psi}\\
        \ddot{\psi}
    \end{pmatrix}}^{\mathcal{X}} 
    &=
    \overbracket{\begin{pmatrix}
        \mathcal{C}_{1}\\
        \mathcal{C}_{2}\\
        0\\
        0\\
        0\\
        \dot{\mathcal{C}}_{1}\\
        \dot{\mathcal{C}}_{2}\\
        0
    \end{pmatrix}}^{\mathcal{C}}
\end{align}

\begin{table}[h!]
    \centering
    \begin{tabular}{ccccc}
     $\mathcal{B}_{11} = \tfrac{1}{4}(2V_{\phi} + P_{\phi}\dot{\phi}^2)\,,$    
     &$\quad$&
     $\mathcal{B}_{14} = - H \mathcal{B}_{17} + \dot{\phi} \mathcal{B}_{12}\,,$ &
     \\
     $4\mathcal{B}_{16} =\mathcal{B}_{36} = 8 \tfrac{k^2}{a^2}\,,$
     &&
     $\mathcal{B}_{31} = 3(2 V_{\phi} - P_{\phi} \dot{\phi}^2)\,,$ &
     \\
     $\mathcal{B}_{52} = (3 H P + P_{\phi}\dot{\phi})\,,$
     &&
     $-6 \mathcal{B}_{12} = 3\mathcal{B}_{21} = -3 P \dot{\phi}\,,$ &
     \\
     $-\tfrac{3}{2} H \mathcal{B}_{17} = H \mathcal{B}_{24} = \mathcal{B}_{27} = 4 - \tfrac{6\kappa^2 \rho}{\tfrac{k^2}{a^2} - 3 \dot{H}}\,,$
     &$\quad$&
     $\mathcal{B}_{34} = -8 \tfrac{k^2}{a^2} + 6 (12H^2 + 8\dot{H} + P \dot{\phi}^2)\,,$ &
     \\
     $\mathcal{B}_{51} = \tfrac{1}{2} \left(2\tfrac{k^2}{a^2} - \tfrac{P_{\phi}}{P} (2 V_{\phi} + P_{\phi}\dot{\phi}^2) + 2 V_{\phi\phi} + P_{\phi\phi}\dot{\phi}^2\right)\,,$
     &&
     $\mathcal{C}_{1} = -\kappa^2 \rho \delta_{m} - \tfrac{3 H}{\tfrac{k^2}{a^2}-3\dot{H}} \dot{\delta}_{m}\,,$
     \\
     $\mathcal{C}_{2} = -\tfrac{2 \kappa^2 \rho}{\tfrac{k^2}{a^2}-3\dot{H}} \dot{\delta}_{m}\,,$ &&
    \end{tabular}
    \captionsetup{labelformat=empty}
    \caption{}
    \label{tab:matrix_terms}
\end{table}
\vskip -5ex
\noindent and solve for $\mathcal{X} = \mathcal{B}^{-1} \mathcal{C}$. By taking the time derivative of Eq.~\eqref{eq:v_equation} and substituting the solutions for each gravitational perturbed variable to obtain the generalized M\'esz\'aros equation of the form 
\begin{align} \label{eq:meszaros_form}
    0 = \ddot{\delta}_{m} + 2 H \nu \dot{\delta}_{m} - \frac{1}{2} \kappa^2 \mu \rho \delta_{m}\,,
\end{align}
where
\begin{align}
    \label{eq:nu} 
    \nu &= 1 + \frac{36 \dot{H} + 3 H P \dot{\phi}^2 + 12  \ddot{H}}{2 H (4 \tfrac{k^2}{a^2} + 3 P \dot{\phi}^2)} \,, \\
    \label{eq:mu} \mu &= 1 + \frac{4 P \dot{\phi}^2}{4 \tfrac{k^2}{a^2} - P \dot{\phi}^2} \,,
\end{align}
and
\begin{align} \label{eq:rho}
    \rho(t) = \frac{3 H_0^2 \Omega_{m0}}{\kappa^2}a^{-3}\,,
\end{align}
during matter epoch where $H_0$ and $\Omega_{m0}$ correspond to the present values of the Hubble and fractional energy density parameters. These equations show the growth of CDM for this action is not explicitly dependent on the potential $V(\phi)$. Regardless, the potential is present in the background equations. Additionally, when consider the case well within the matter dominated era, where $a \propto t^{\tfrac{2}{3}}$, the derivatives of the Hubble parameter can be taken to be $\dot{H} = -\tfrac{3}{2} H^2$ and $\ddot{H} = \tfrac{9}{2} H^3$ such that the $\nu$ coefficients given by Eq.~\eqref{eq:nu} simplifies to $\nu = 1 + \frac{3 P \dot{\phi}^2}{2(4 \tfrac{k^2}{a^2} + 3 P \dot{\phi}^2)}$ such that as $P \rightarrow 0$, the GR limit is obtained where $\mu = \nu = 1$.

Moving forward, we will adopt the scale factor $a$ to describe cosmic evolution such that such that primes ($'$) represent derivatives with respect to $a$, which results in
\begin{align} \label{eq:meszaros_in_a}
    0 = -\frac{3}{2} \Omega_{m}(a) \bar{\mu} \delta_{m}(a) + a \left(2 \bar{\nu} + 1 + a \frac{h'(a)}{h(a)}\right) \delta_{m}'(a) + a^2 \delta_{m}''(a),
\end{align}
where $\bar{\nu}$ and $\bar{\mu}$ correspond to $\nu$ (Eq.~\eqref{eq:nu}) and $\mu$ (Eq.~\eqref{eq:mu}) expressed in terms of the scale factor, $h$ is the ratio of the Hubble parameter to its present value and 
\begin{align} \label{eq:matter_density}
    \Omega_{m}(a) = \frac{\Omega_{m0} a^{-3}}{h(a)^2}\,.
\end{align}
Moreover, the variable $D$ is used to describe the ratio of perturbation amplitude at an arbitrary scale factor $a$ and its value at initial scale factor $a_{i}$
\begin{align}\label{eq:growth_D}
    D(a) = \frac{\delta_{m}(a)}{\delta_{m}(a_{i})}\,,
\end{align}
Therefore, the dynamics of matter perturbation growth are described by Eqs~(\ref{eq:meszaros_in_a}-\ref{eq:growth_D}). Additionally, the growth index can also be obtained from these equations. Another definition for growth can be given as~\cite{Linder:2005in,Pouri:2014nta}
\begin{align}\label{eq:growth_D_redefinition}
    D(a) = \exp{\left(\int_{1}^{a} \frac{\Omega_{m}(\tilde{a})^{\gamma(\tilde{a})}}{\tilde{a}}\, d\tilde{a}\right)}\,,
\end{align}
where at present $a = 1$ and $\gamma(a)$ is the varying growth index. Hence, the growth rate clustering is given by~\cite{Pouri:2014nta}
\begin{align}\label{eq:growth_rate}
    g(a) = \frac{d \ln \delta_{m}}{d \ln a} \simeq \Omega_{m}(a)^{\gamma(a)}\,. 
\end{align}
Substituting Eq.~\eqref{eq:growth_rate} in Eq.~\eqref{eq:meszaros_in_a} yields to
\begin{align}\label{eq:growth_index_in_a}
    0 = -\frac{3}{2} \bar{\mu} \Omega_m(a) + \Omega_{m}(a)^{\gamma(a)} + \Omega_{m}(a) \left(a (1-2\gamma(a))\frac{h'(a)}{h(a)} + 2 \bar{\nu} - 3 \gamma(a) + a \gamma'(a) \ln\Omega_{m}(a) \right)\,.
\end{align}
The growth factor is set to have a linear parametrization~\cite{Linder:2005in,Wu:2009zy,Polarski:2007rr,Dossett:2010gq}
\begin{align}\label{eq:dynamical_growth_index}
    \gamma(a) = \gamma_{0} + (1-a) \gamma_1\,,
\end{align}
where $\gamma_{0}$ and $\gamma_{1}$ are constants. At current time $\gamma(1) = \gamma_{0}$. Setting Eq.~\eqref{eq:growth_index_in_a} at present time
\begin{align} \label{eq:growth_index_current}
    0 = -\frac{3}{2} \bar{\mu}^{(a=1)} \Omega_{m0} + \Omega_{m0}^{2\gamma_0} + \Omega_{m0}^{\gamma_0} \left(-3 \gamma_0 + 2 \bar{\nu}^{(a=1)} - \gamma_1 \ln \Omega_{m0} + (1-2\gamma_0)h'(1) \right)\,,
\end{align}
the growth index value can be obtained. The result for a constant growth index can be determined when $\gamma_1 = 0$ in the above equation.

\section{Models}\label{sec:models}

In the previous section, we have seen that the results dependent on the matter perturbations are independent of the potential $V(\phi)$ introduced in the action Eq.~\eqref{eq:action}. While this statement holds for Eqs.~(\ref{eq:meszaros_form},\ref{eq:growth_index_current}), the potential has an impact on the background equations~(\ref{eq:field_equations_background_W00}-\ref{eq:field_equations_background_Wphi}). By considering different models, we will illustrate the growth factor and growth index behaviour when considering $k$-dependencies in contrast with the result obtained when applying the subhorizon limit $k \gg aH$. Hence, we can obtain the visual representation of whether the application of the subhorizon limit from the start is an adequate assumption for these particular models. Note, $H_{0} = 67.4\text{ km }\text{s}^{-1}\text{ Mps}^{-1}$  and $\Omega_{m0} = 0.315$ parameters are set using data from Planck collaboration~\cite{Planck:2018vyg} for illustrative purposes.

In general, we will consider cases where $P = 1$. This implies that when looking at cases when $a = 1$, where $h(1) = 1$, $\phi_0=\phi(1)$ and $\phi'(1)$ are set boundary conditions, would yield an identical result for Eq.~\eqref{eq:growth_index_current} regardless of the potential considered, unless $P(\phi)$ function is varied. It should be noted that changing $\phi'(1)$ would result in slightly different results, but when imposing the equation of state~\eqref{eq:effective_EoS} $\omega_{\text{eff}} = -1$ at current time, results in $\phi'(1) = 0$. This can be seen by considering the Friedmann equation~\eqref{eq:field_equations_background_W00} in terms of $a$ (instead of $t$) at $a=1$
\begin{align}\label{eq:V(1)}
    V(a=1) = 6H_{0}^{2}(1-\Omega_{m0}) - \frac{1}{2} H_{0}^{2} \phi'(1)^2\,,
\end{align}
which provides an expression for $V(a=1)$. Substituting in the equation of state~\eqref{eq:effective_EoS} (transformed to be $a$ dependent) gives
\begin{align} \label{eq:EoS_Cases}
    \omega_{\text{eff}}(1) = -1 + \frac{ \phi'(1)^2 }{6(1-\Omega_{m0})}\,,
\end{align}
for which $\phi'(1) = 0$ is required to attain a dark energy equation of state. 

\begin{figure}[h!]
    \centering
    \includegraphics[width=0.8\textwidth]{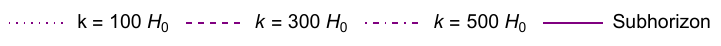}
    \vskip 2mm
    \includegraphics[width=\textwidth]{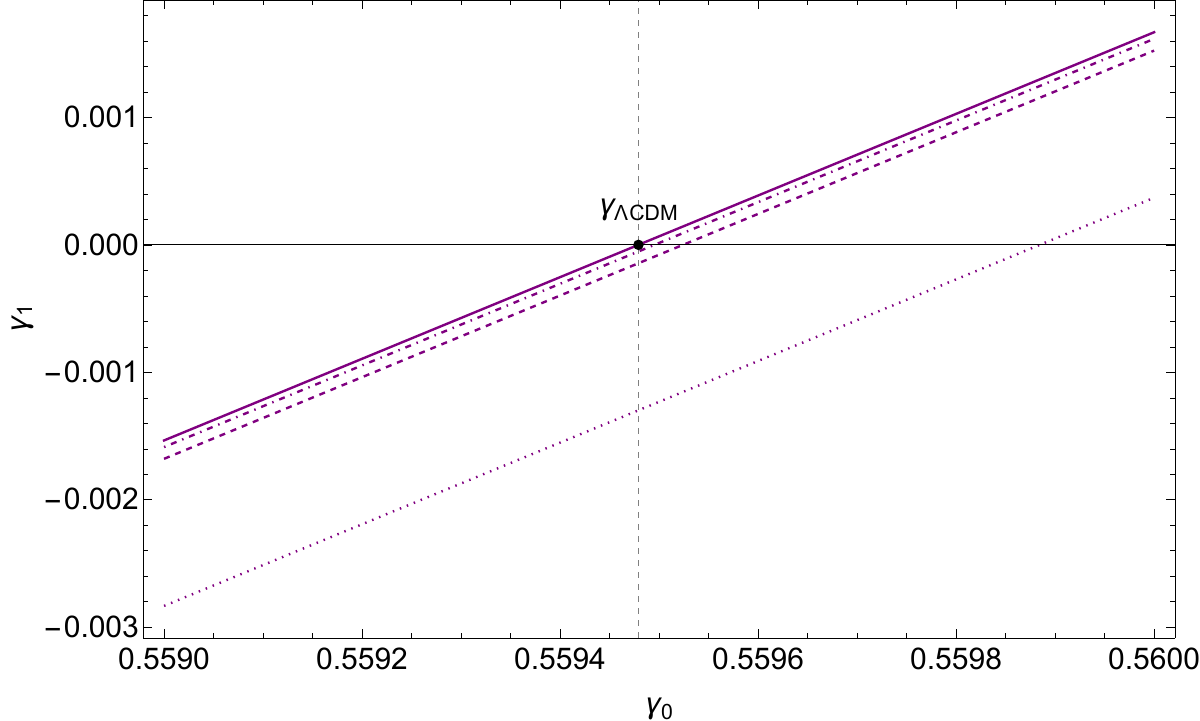}
    \caption{Growth index plots for cases of the form $-T - f(\phi) + X$ comparing $\gamma_1$ to $\gamma_0$. The $\Lambda$CDM limits $\gamma_{\Lambda\text{CDM}} = 0.5594792$ is can be obtained using different $k$ values as depicted with the vertical dashed line.}
    \label{fig:GrowthIndex_Case1to3}
\end{figure}

Regardless of different background equations governed by different potentials, it will not alter the main result. Fig.~\ref{fig:GrowthIndex_Case1to3} plots the dynamical growth index at $a= 1$, illustrating the balance between $\gamma_0$ and $\gamma_1$ to obtain the $\Lambda$CDM values, $\gamma_{\Lambda\text{CDM}}$. The plot shows that as $k$ increases, the results approach that obtained for the subhorizon limit. In Ref.~\cite{Capozziello:2023giq}, in the case of $f(T)$ teleparallel gravity, it was shown for the majority of the cases considered, $k > 100 H_{0}$ typically yields results well within the subhorizon limit even when considering $\gamma_0$ value up to 4 significant figures. Notice, the result presented here implies $k$-depedency. In Table~\ref{tab:Case1to3_gamma}, the values for $(\gamma_0,\gamma_1)$ have been summarised for different $k$ values and at the subhorizon limit at which the $\Lambda$CDM result it attained, and at $\gamma_{0}=0.55$ and $\gamma_0 = 0.56$, which are commonly quoted values.

\begin{table}[h!]
    \centering
    \begin{tabular}{|c|>{\centering\arraybackslash}p{2cm}|>{\centering\arraybackslash}p{4cm}|>{\centering\arraybackslash}p{3cm}|>{\centering\arraybackslash}p{3cm}|}
    \hline
    \multirow{2}{*}{} & \multicolumn{4}{c|}{$(\gamma_0,\gamma_1)$}\\
    \cline{2-5}
    & $\gamma = \gamma_0$ & $\Lambda$CDM solution at $\gamma = \gamma_0$ & $\gamma_0 = 0.55$ & $\gamma_0 = 0.56$\\
    \hline
    $k = 100 H_0$ & $(0.559884,0)$ & $(0.5594792, -0.001299)$ & $(0.55,-0.031657)$ & $(0.56, 0.000370)$\\
    $k = 300 H_0$ & $(0.559524,0)$ & $(0.5594792, -0.000144)$ & $(0.55,-0.030503)$ & $(0.56, 0.001525)$\\
    $k = 500 H_0$ & $(0.559495,0)$ &  $(0.5594792, -0.000052)$ & $(0.55,-0.03041)$ & $(0.56, 0.001617)$\\
    Subhorizon & $(0.559479,0)$ & $(0.5594792, 2.8 \times 10^{-16})$ & $(0.55,-0.030358)$ & $(0.56, 0.001669)$\\
    \hline
    \end{tabular}
    \caption{Growth index values satisfying the dynamical relationship $\gamma = \gamma_0 + (1-a) \gamma_1$ at different $k$ values and at subhorizon limit corresponding to the model $R + V(\phi) + X$ depicted in Fig.~\ref{fig:GrowthIndex_Case1to3}. The second column represents the value at which the growth index is taken to be a constant such that $\gamma_1 = 0$. The third column corresponds to a $\gamma$ value at which the growth index is constant at the $\Lambda$CDM limit. The fourth column gives the values to obtain $\gamma_0 = 0.55$. The last column gives the values to obtain $\gamma_0 = 0.56$.}
    \label{tab:Case1to3_gamma}
\end{table}

\subsection{\texorpdfstring{$\mathcal{L} = R + V_{0} \phi^2 + X$}{}}\label{sec:Case1}

In the first case, we consider the subcase of potential $V = V_{0} \phi(a)^2$~\cite{Felder:2002jk} and the kinetic term $P=1$ in Eq.~\eqref{eq:action}, where $V_{0}$ is a constant. Substituting within the background Friedmann equation~\eqref{eq:field_equations_background_W00} gives
\begin{align} \label{eq:Case1_Friedmann}
    h(a)^{2} = \frac{2 (6 H_{0}^2 \Omega_{m0} + V_{0} a^3 \phi(a)^{2})}{a^3 H_{0}^2 (12 - a^2 \phi'(a)^2)}\,.
\end{align}
To determine the value of $V_{0}$, Friedmann equation is evaluated at current time using Eq.~\eqref{eq:V(1)} and imposing the condition $\phi'(1) = 0$:
\begin{align} \label{eq:Case1_V0}
    V_{0} = \frac{4H_{0}^2(1-\Omega_{m0})}{\phi_{0}^{2}}\,,
\end{align}
where $\phi_{0} = \phi(1)$ is an unknown constant.

\begin{figure}[h!]
    \centering
    \includegraphics[width=0.7\textwidth]{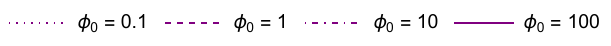}
    \vskip 2mm
    \includegraphics[width=0.7\textwidth]{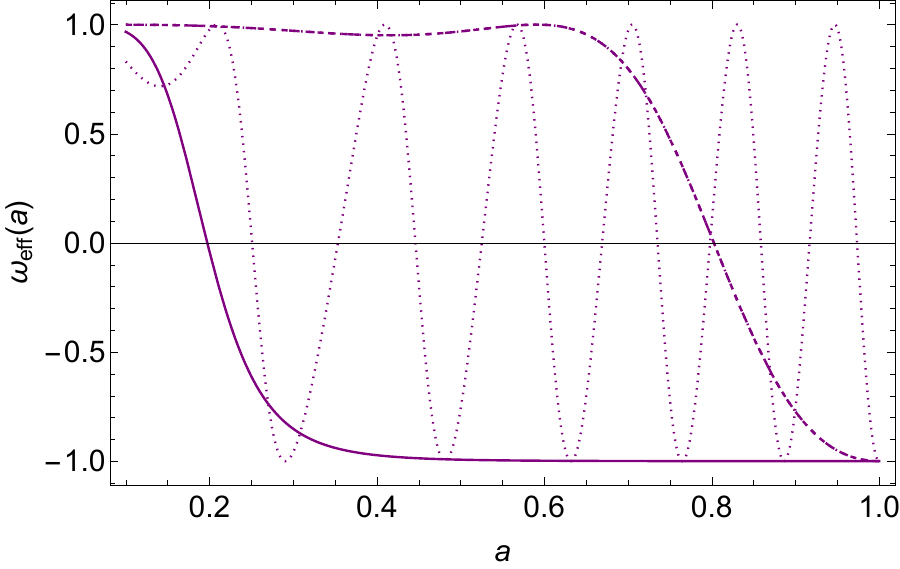}
    \caption{Effective equation of state $\omega_{\text{eff}}$ for different values of $\phi_{0}$ for the case $V(\phi) = V_0 \phi^2$. Cases $\phi_{0} = 1$ and $\phi_{0} = 10$ overlap each other.}
    \label{fig:omega_effective_Case1}
\end{figure}

Next, the scalar field equation~\eqref{eq:field_equations_background_Wphi} is obtained by substituting Eq.~\eqref{eq:Case1_V0} and $\phi'(1) = 0$:
\begin{align} \label{eq:Case1_scalar_equation}
    0 = 12 (1-\Omega_{m0}) \frac{\phi(a)}{\phi_{0}^2} + a h(a) ((4 h(a) + a h'(a)) \phi'(a) +a h(a) \phi''(a))\,.
\end{align}
Eq.~\eqref{eq:Case1_Friedmann} and Eq.~\eqref{eq:Case1_scalar_equation} are solved simultaneously for $h$ and $\phi$, given a choice of $\phi_0$. Therefore, the equation of state for different $\phi_0$ values can be seen in Fig.~\ref{fig:omega_effective_Case1}. For $\phi_{0} = 0.1$ the behaviour is seen to be oscillatory with the period decreasing as late-times are approached. For the cases $\phi_{0} = 1$ and $\phi_{0} = 10$ (overlapping) exhibit a stiff behaviour $p_{\text{eff}}
 ~ \rho_{\text{eff}}$ before approaching $\omega_{\text{eff}} = -1$ (a given condition). At $\phi_{0}=100$, a more dynamical behaviour for the effective equation of states is given, which is favoured over oscillatory and stiff plots.

\begin{figure}[h!]
    \centering
    \begin{minipage}{\textwidth}
    \centering
    \includegraphics[width=0.7\textwidth]{Plots/legend1DeltaD.pdf}
    \end{minipage}
    \subfloat[$\phi_0 = 0.1$]{\includegraphics[width=0.45\textwidth]{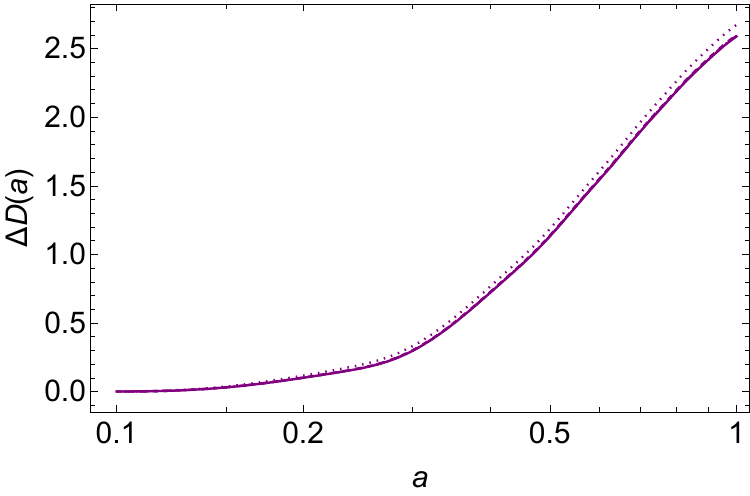}\label{fig:GrowthFactor_Case1_0.1}}
    \qquad
    \subfloat[$\phi_0 = 1$]{\includegraphics[width=0.45\textwidth]{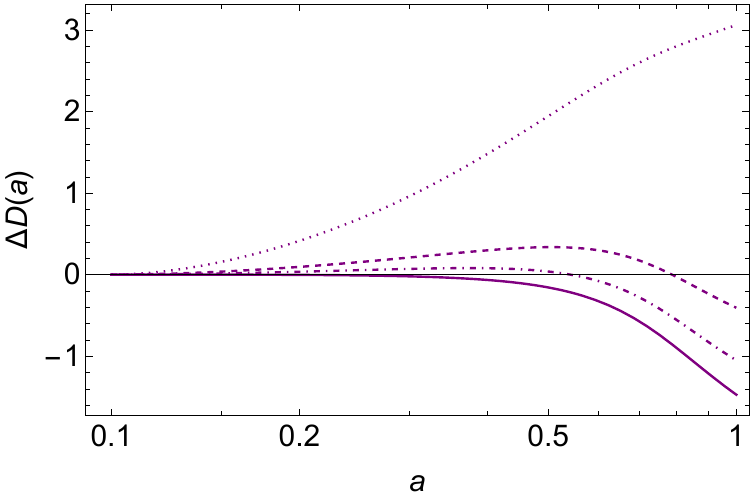}\label{fig:GrowthFactor_Case1_1}}\\
    \subfloat[$\phi_0 = 10$]{\includegraphics[width=0.45\textwidth]{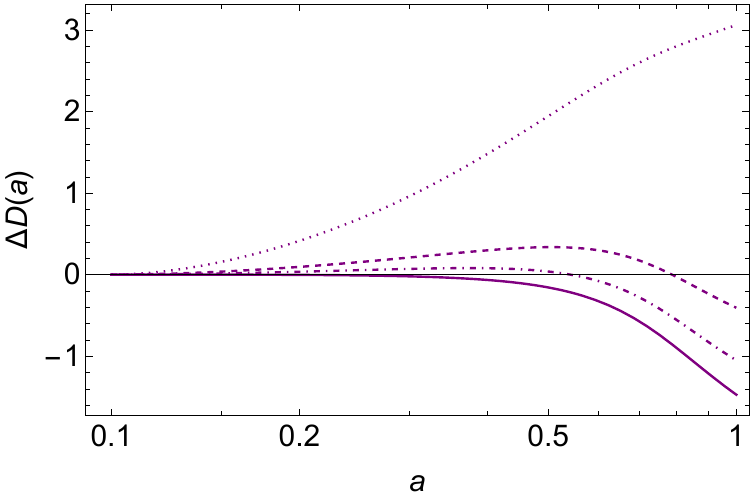}\label{fig:GrowthFactor_Case1_10}}
    \qquad
    \subfloat[$\phi_0 = 100$]{\includegraphics[width=0.45\textwidth]{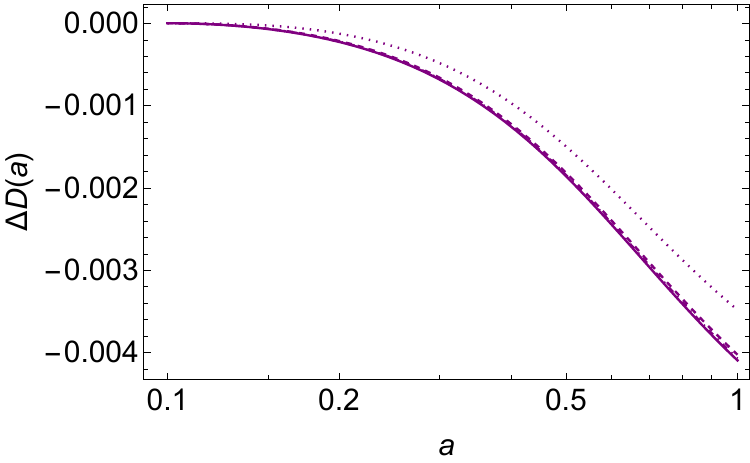}\label{fig:GrowthFactor_Case1_100}}
    \caption{Difference in growth factor from the $\Lambda$CDM solution for the case $V(\phi) = V_{0}\phi^{2}$. Each quadrant presents the solutions for set boundary condition for $\phi_{0}$. Each plot compares values of different $k$ values and the subhorizon limit result obtained through the immediate assumption of $k \gg aH$.}
    \label{fig:Growth_Factor_Case1}
\end{figure}

The growth factor for this model is given by Fig.~\ref{fig:Growth_Factor_Case1}. Each quadrant depicts plots for different $\phi_{0}$ values, showing the deviation of growth factor solutions for different $k$ values and subhorizon result from the $\Lambda$CDM solution. It should be noted that for large values such as $\phi_{0} = 100$, for a range of $k$ values, the solutions approach that of the subhorizon and $\Lambda$CDM limit as given by Fig.~\ref{fig:GrowthFactor_Case1_100}. For $\phi_{0} = 0.1$ in Fig.~\ref{fig:GrowthFactor_Case1_0.1}, deviations from $\Lambda$CDM are more significant while for $k > 100 H_0$ results coincide with that of the subhorizon limit. Additionally, $\phi_0 = 1$ (Fig.~\ref{fig:GrowthFactor_Case1_1}) and $\phi_0 = 10$ (Fig.~\ref{fig:GrowthFactor_Case1_10}) have identical solutions, as predicted by the behaviour of $\omega_{\text{eff}}$ in Fig.~\ref{fig:omega_effective_Case1}. As seen in the case of $f(T)$ gravity~\cite{Capozziello:2023giq}, typically, results beyond $k= 300 H_{0}$ are well within the subhorizon regime and would not appear in the plots.

\subsection{\texorpdfstring{$\mathcal{L} = R + V_{0} \phi^4 + X$}{}}\label{sec:Case2}

Next, we will consider a case similar to the case in Sec.~\ref{sec:Case1}, this time with the potential of the form $V(\phi) = V_0 \phi^4$~\cite{Linde:1983gd} and $P = 1$. Substituting this Lagrangian into the first Friedmann equation~\eqref{eq:field_equations_background_W00} to obtain
\begin{align}\label{eq:Case2_Friedmann}
    h(a)^2 = \frac{2(6 H_0^2 \Omega_{m0} + a^3 V_{0}\phi(a)^4)}{a^3 H_0^2 (12 - a^2 \phi'(a)^2)}\,.
\end{align}
Once again, the constant $V_{0}$ is obtained by setting the above equation at current time, represented by $a = 1$ such that
\begin{align}\label{eq:Case2_V0}
    V_{0} = \frac{6H_{0}^2 (1-\Omega_{m0})}{\phi_{0}^4}\,.
\end{align}

\begin{figure}[h!]
    \centering
    \includegraphics[width=0.7\textwidth]{Plots/legend2_phi.pdf}
    \vskip 2mm
    \includegraphics[width=0.7\textwidth]{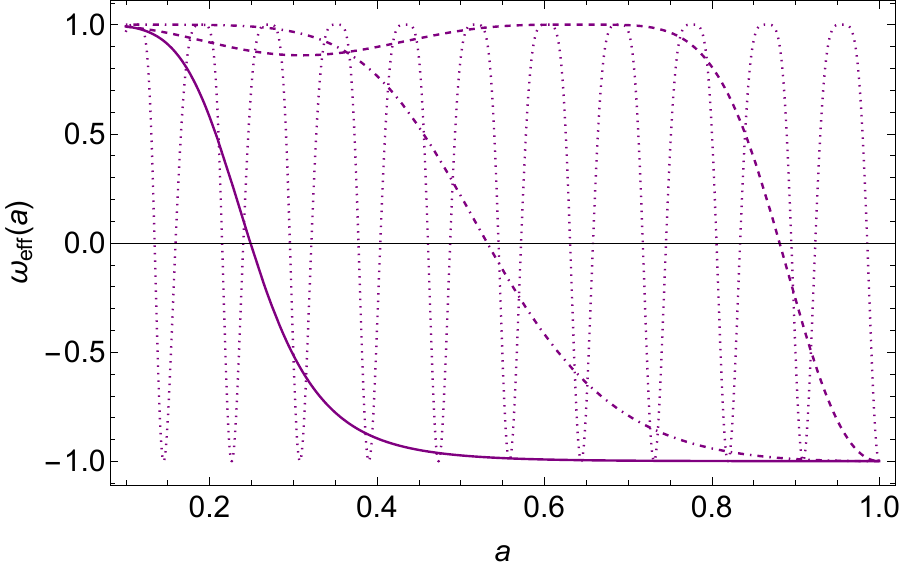}
    \caption{Effective equation of state $\omega_{\text{eff}}$ for different values of $\phi_{0}$ for the case $V(\phi) = V_0 \phi^4$.}
    \label{fig:omega_effective_Case2}
\end{figure}

\begin{figure}[h!]
    \centering
    \begin{minipage}{\textwidth}
    \centering
    \includegraphics[width=0.7\textwidth]{Plots/legend1DeltaD.pdf}
    \end{minipage}
    \subfloat[$\phi_0 = 0.1$]{\includegraphics[width=0.45\textwidth]{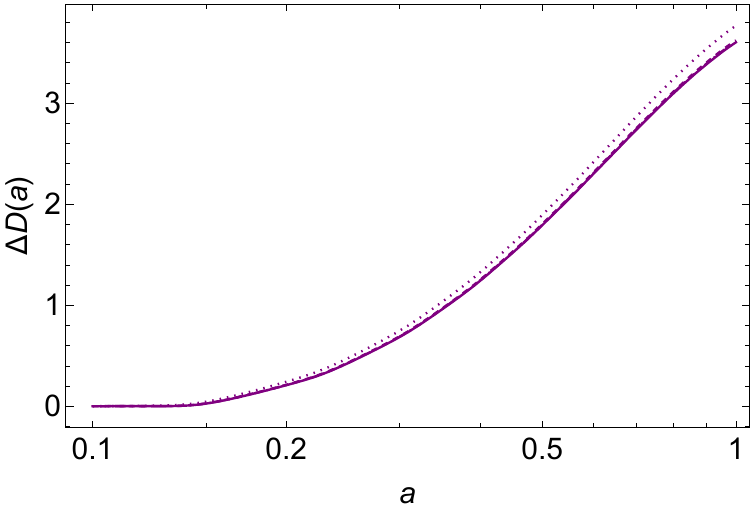}\label{fig:GrowthFactor_Case2_0.1}}
    \qquad
    \subfloat[$\phi_0 = 1$]{\includegraphics[width=0.45\textwidth]{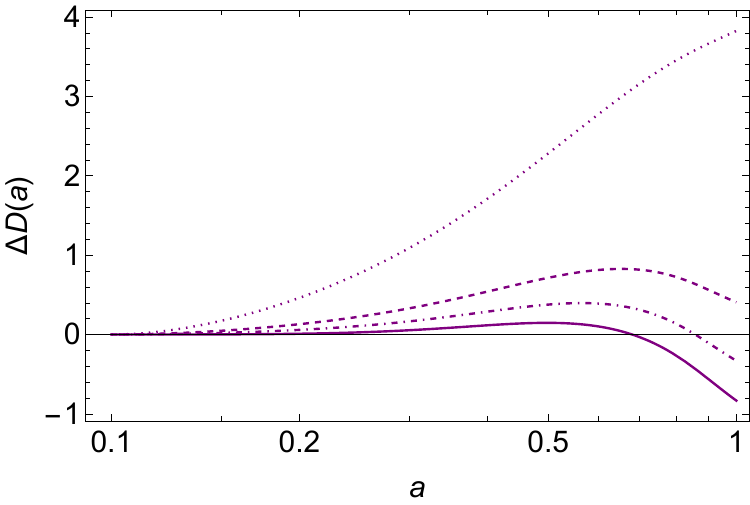}\label{fig:GrowthFactor_Case2_1}}\\
    \subfloat[$\phi_0 = 10$]{\includegraphics[width=0.45\textwidth]{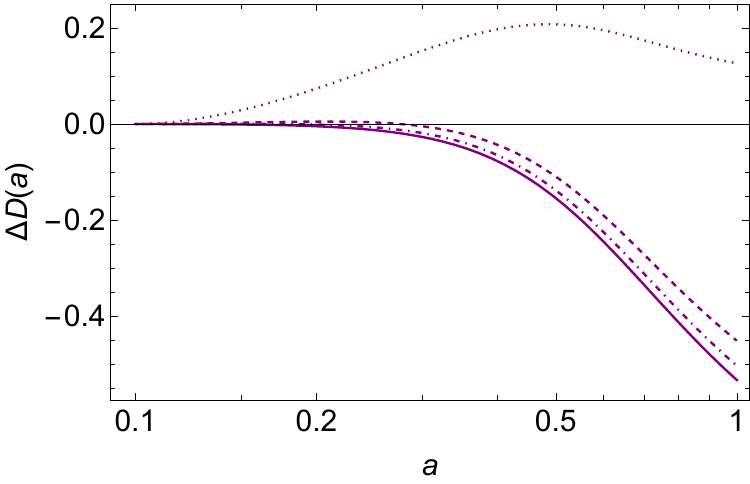}\label{fig:GrowthFactor_Case2_10}}
    \qquad
    \subfloat[$\phi_0 = 100$]{\includegraphics[width=0.45\textwidth]{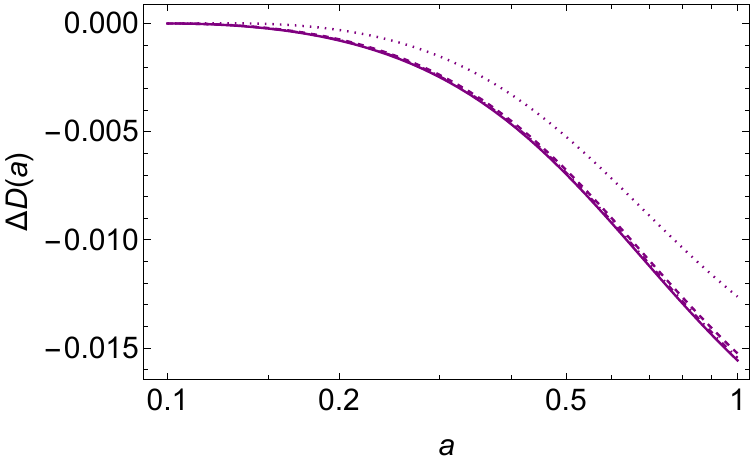}\label{fig:GrowthFactor_Case2_100}}
    \caption{Difference in growth factor from the $\Lambda$CDM solution for the case $V(\phi) = V_{0}\phi^{4}$. Each quadrant presents the solutions for set boundary condition for $\phi_{0}$. Each plot compares values of different $k$ values and the subhorizon limit result obtained through the immediate assumption of $k \gg aH$.}
    \label{fig:Growth_Factor_Case2}
\end{figure}

Additionally, the scalar field equation~\eqref{eq:field_equations_background_Wphi} can be expressed as
\begin{align}
    \label{eq:Case2_Scalar}
    0 = -12(\Omega_{m0}-1)\frac{\phi(a)^3}{\phi_{0}^4} + a h(a) ((4h(a) + ah'(a))\phi'(a) +a h(a) \phi''(a))\,.
\end{align}
The solutions for $h$ and $\phi$ can be obtained at the background level provided a choice of $\phi_{0}$ value is assumed. Hence, the equation of state described by Eq.~\eqref{eq:effective_EoS} for this case is depicted in Fig.~\ref{fig:omega_effective_Case2}. Once again, the case for $\phi_0 = 0.1$ is oscillatory, but with a shorter period when compared to the previous case in Sec.~\ref{sec:Case1}. This coincides with the fact that the potential has a higher order index. Due to this quadruple order, this time only the case $\phi_0 = 1$ results in stiff equation of state while $\phi_0 = 10$ and $\phi_0 = 100$ give more of a dynamical behaviour, yielding a more viable result. Oscillatory effective equation of state is not favoured due to the implication that matter dominated era experiences periods of accelerated and decelerated expansion, resulting in a different large scale structure observed today. In fact, this is reflected further in the overall significant slower growth than the $\Lambda$CDM limit for $\phi_{0} = 0.1$ as seen in Fig.~\ref{fig:GrowthFactor_Case2_0.1}.

The general M\'esz\'aros equation~\eqref{eq:meszaros_in_a} is evaluated for different values of $\phi_0$ for a variety of $k$ values to evaluate whether $k$-dependency is present for this model. These results are represented in Fig.~\ref{fig:Growth_Factor_Case2}. These results are similar in nature to those of analyzed in Fig.~\ref{fig:Growth_Factor_Case1}, apart from $\phi_0 = 10$ in Fig.~\ref{fig:GrowthFactor_Case2_10}. The overall difference is that all results are lower, representing a slower growth, than their quadratic potential counterparts in Sec.~\ref{sec:Case1}.

\subsection{\texorpdfstring{$\mathcal{L} = R + V_{0} e^{\lambda \phi} + X$}{}}\label{sec:Case3}

\begin{figure}[h!]
    \centering
    \includegraphics[width=0.7\textwidth]{Plots/legend2_phi.pdf}
    \vskip 2mm
    \includegraphics[width=0.7\textwidth]{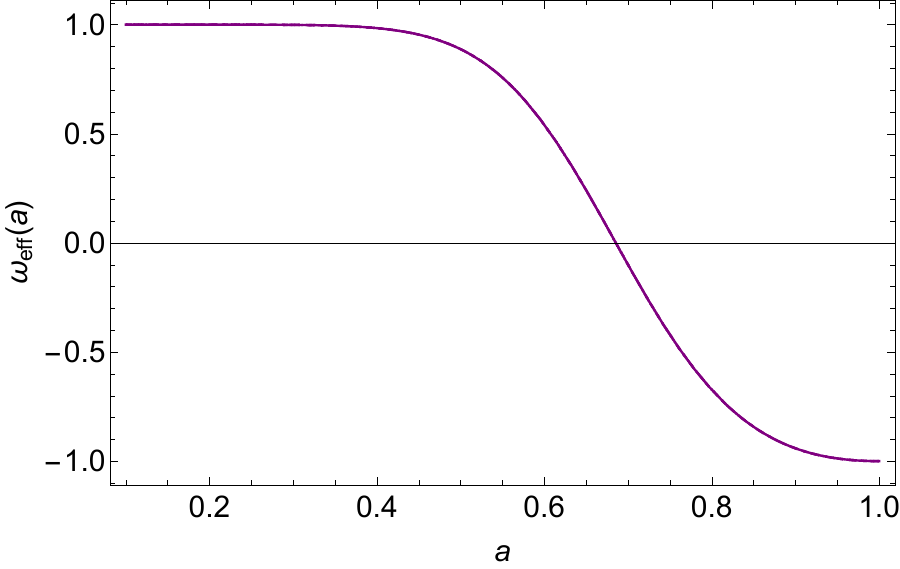}
    \caption{Effective equation of state $\omega_{\text{eff}}$ for different values of $\phi_{0}$ for the case $V(\phi) = V_0 e^{\lambda \phi}$ when $\lambda = 1$. All cases of $\phi_{0}$ considered overlap.}
    \label{fig:omega_effective_Case3}
\end{figure}

In Eq.~\eqref{eq:action}, the action is rewritten by setting $V = V_0 e^{\lambda\phi}$~\cite{Copeland:1997et} where $V_{0}$ and $\lambda$ are constants, and $P = 1$. Our system of background equations is given by the first Friedmann equation~\eqref{eq:field_equations_background_W00} and Klein-Gordon scalar field equation~\eqref{eq:field_equations_background_Wphi}, which for this model is expressed as
\begin{align}
    \label{eq:Case3_Friedmann} h(a)^2 &= \frac{2( V_0 a^3 e^{\lambda \phi(a)} + 6 H_0^2 \Omega_{m0})}{a^3 H_{0}^2 (12 - a^2 \phi'(a)^2)}\,,\\
    \label{eq:Case3_ScalarEq} 0 &= V_0 \lambda e^{\lambda \phi(a)} + a H_{0}^2 h(a) ((4 h(a) + ah'(a))\phi'(a) + a h(a) \phi''(a))\,.
\end{align}
An expression for $V_{0}$ is obtained by setting Eq.~\eqref{eq:Case3_Friedmann} at $a = 1$ to obtain
\begin{align}\label{eq:Case3_V0}
    V_{0} = 6 H_{0}^2 e^{-\lambda\phi_0}(1-\Omega_{m0})\,,
\end{align}
where $\phi_{0}$ will be a constant varied for different subcases and $\phi'(1) = 0$ is set. At first, we will consider the case where $\lambda = 1$ in order to see the behaviour of the growth factor for different $k$ values. Thus, the numerical solutions for $h$ and $\phi$ were determined. Thus, the behaviour of the effective equation of state is illustrated in Fig.~\ref{fig:omega_effective_Case3}, where all values of $\phi_{0}$ considered give a seemingly identical result. It should be noted that some differences are in fact present at small orders of magnitude ($\lesssim |10^{-8}|$) throughout the scale considered.

\begin{figure}[h!]
    \centering
    \begin{minipage}{\textwidth}
    \centering
    \includegraphics[width=0.7\textwidth]{Plots/legend1DeltaD.pdf}
    \end{minipage}
    \subfloat[$\phi_0 = 0.1$]{\includegraphics[width=0.45\textwidth]{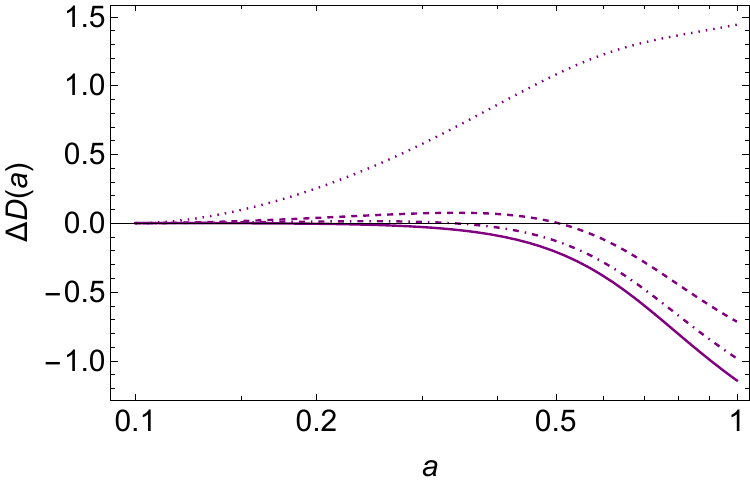}\label{fig:GrowthFactor_Case3_0.1}}
    \qquad
    \subfloat[$\phi_0 = 1$]{\includegraphics[width=0.45\textwidth]{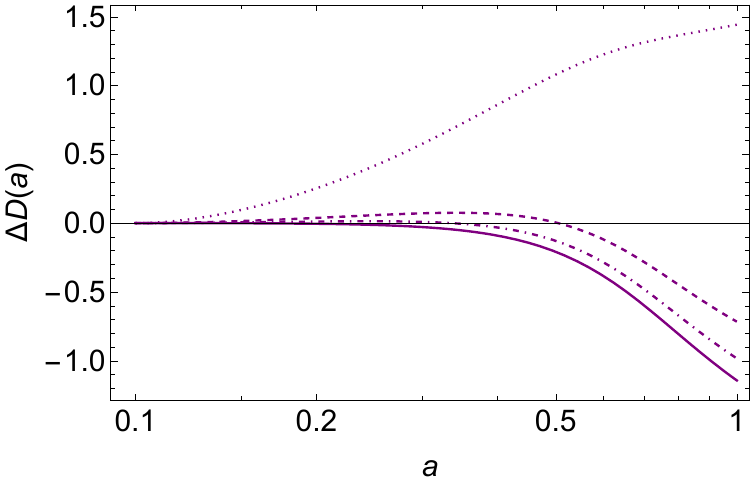}\label{fig:GrowthFactor_Case3_1}}\\
    \subfloat[$\phi_0 = 10$]{\includegraphics[width=0.45\textwidth]{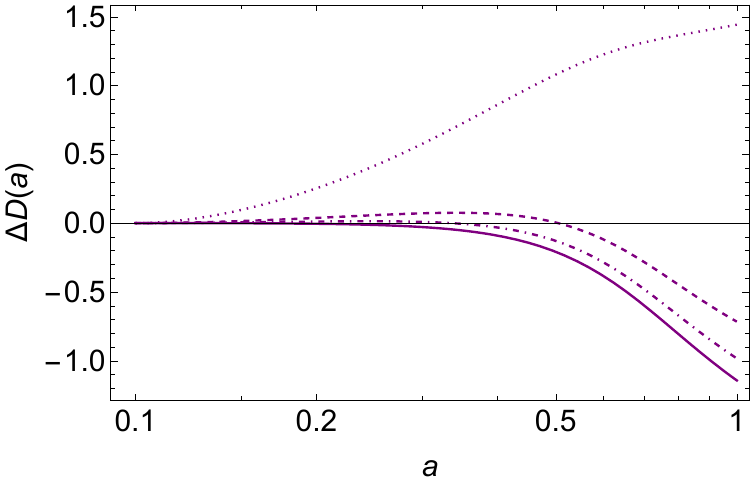}\label{fig:GrowthFactor_Case3_10}}
    \qquad
    \subfloat[$\phi_0 = 100$]{\includegraphics[width=0.45\textwidth]{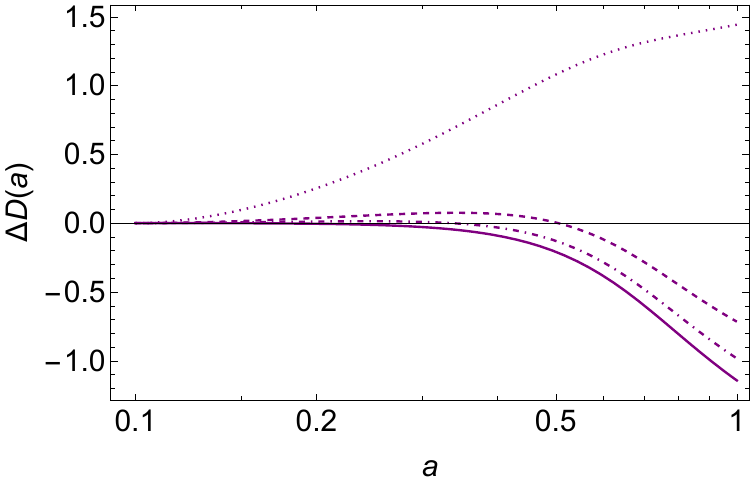}\label{fig:GrowthFactor_Case3_100}}
    \caption{Difference in growth factor from the $\Lambda$CDM solution for the case $V(\phi) = V_{0} e^{\lambda \phi}$ when $\lambda = 1$. Each quadrant presents the solutions for set boundary condition for $\phi_{0}$. Each plot compares values of different $k$ values and the subhorizon limit result obtained through the immediate assumption of $k \gg aH$.}
    \label{fig:Growth_Factor_Case3}
\end{figure}

The difference in growth factor to the $\Lambda$CDM solution is illustrated in Fig.~\ref{fig:Growth_Factor_Case3}. While the plots depict $k$ dependency similar to the cases in Figs~(\ref{fig:GrowthFactor_Case1_1},\ref{fig:GrowthFactor_Case2_1}), quadrants~(\ref{fig:GrowthFactor_Case3_0.1}-\ref{fig:GrowthFactor_Case3_100}) appear to have identical behaviour: for $k = 100 H_{0}$ the growth is slower than the $\Lambda$CDM limit. For higher values of $k$ such as $k = 300 H_{0}$ and $k = 500 H_{0}$ the results are higher than the $\Lambda$CDM limit, with the subhorizon solution giving the fastest growth within the subcase. Since the dependency on $\phi$ appears within the exponential function, the differences in plots can be seen when considering the last $10^{-6}$ of the scale.

\begin{figure}[h!]
    \centering
    \includegraphics[width=0.7\textwidth]{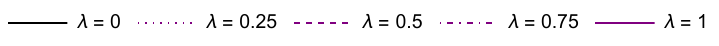}
    \vskip 2mm
    \includegraphics[width=0.7\textwidth]{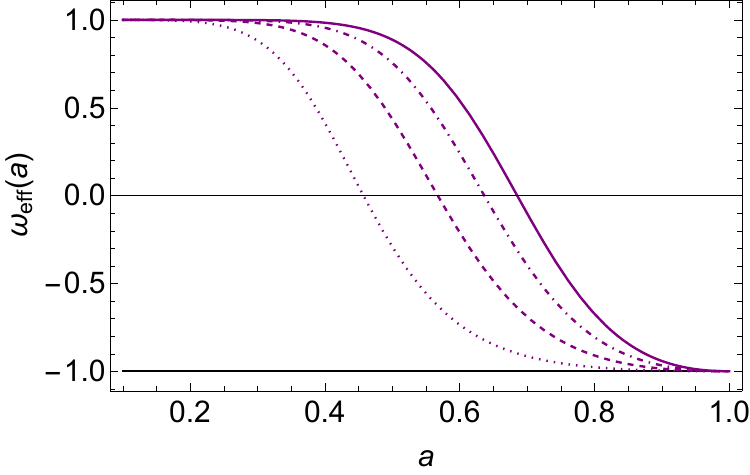}
    \caption{Effective equation of state $\omega_{\text{eff}}$ for different values of $\lambda$ for the case $V(\phi) = V_0 e^{\lambda \phi}$ when $\phi_{0} = 1$ and $\phi'(1) = 0$.}
    \label{fig:omega_effective_Case4}
\end{figure}

Finally, we also conduct a comparison when different $\lambda$ values are considered. It should be noted that $|\lambda|$ is considered as negative values yield to the same results since the sign present in Eqs~(\ref{eq:Case3_Friedmann}-\ref{eq:Case3_ScalarEq}) is accounted for by the constant $V_{0}$ in Eq.~\eqref{eq:Case3_V0}. Additionally, due to minute differences in Fig.~\ref{fig:Growth_Factor_Case3}, we will only consider the background boundary $\phi_{0} = 1$ only. The effective equation of state~\eqref{eq:EoS_Cases} for different $\lambda$ values is illustrated in Fig.~\ref{fig:omega_effective_Case4}. $\lambda = 0$ yields to a constant value of $\omega_{\text{eff}} = -1$. Albeit the result is identical to that of GR, it should be noted that the action is $\mathcal{R} + V_{0} + X$ for $\lambda = 0$, making the model distinguishable from $\Lambda$CDM. Additionally, we only considered values in the range $\lambda \in [0,1]$, since larger values of $\lambda$ start to give a stiff-like behaviour.

\begin{figure}[h!]
    \centering
    \begin{minipage}{\textwidth}
    \centering
    \includegraphics[width=0.7\textwidth]{Plots/legend4_lambda.pdf}
    \end{minipage}
    \subfloat[$k=100H_0$]{\includegraphics[width=0.45\textwidth]{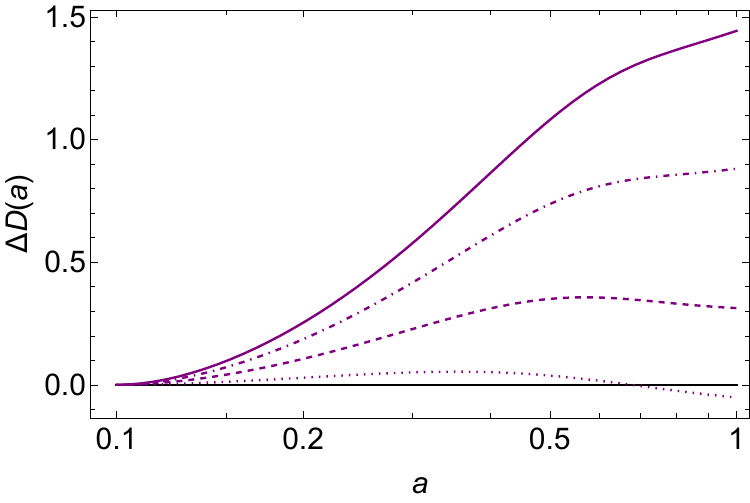}\label{fig:GrowthFactor_Case4_100}}
    \qquad
    \subfloat[$k =300H_0 $]{\includegraphics[width=0.45\textwidth]{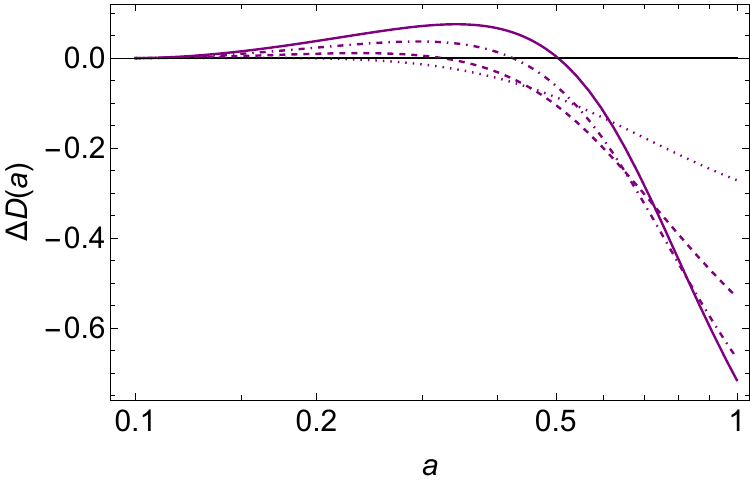}\label{fig:GrowthFactor_Case4_300}}\\
    \subfloat[$k = 500 H_0$]{\includegraphics[width=0.45\textwidth]{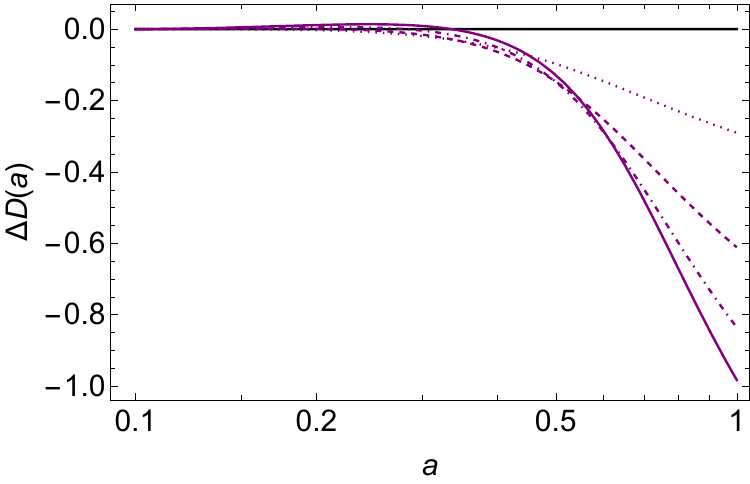}\label{fig:GrowthFactor_Case4_500}}
    \qquad
    \subfloat[Subhorizon]{\includegraphics[width=0.45\textwidth]{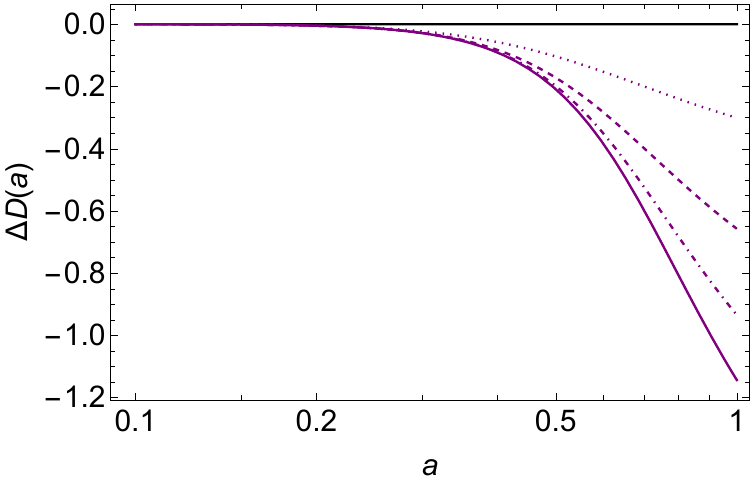}\label{fig:GrowthFactor_Case4_Subhorizon}}
    \caption{Difference in growth factor from the $\Lambda$CDM solution for the case $V(\phi) = V_{0} e^{\lambda \phi}$ when $\lambda = 1$. Each quadrant presents the solutions for set boundary condition for $\phi_{0}$. Each plot compares values of different $k$ values and the subhorizon limit result obtained through the immediate assumption of $k \gg aH$.}
    \label{fig:Growth_Factor_Case4}
\end{figure}

The growth factor plots comparing different $\lambda$ values, split up for different $k$ values, is given by Fig.~\ref{fig:Growth_Factor_Case4}. Once again, in the case of $\lambda = 0$, we retrieve a result analogous to the $\Lambda$CDM limit. In the case of $k = 100 H_0$ in Fig.~\ref{fig:GrowthFactor_Case4_100} shows an increasing growth factor as the $\lambda$ value increases. This is in contrast with the rest of quadrants (Fig.~(\ref{fig:GrowthFactor_Case4_300}-\ref{fig:GrowthFactor_Case4_Subhorizon})), where the growth is seen to decrease as $\lambda$ increases. Additionally, solutions for the latter cases are closer to each other for different $\lambda$ values.

\section{Conclusion}\label{sec:conclu}

The evolution profile of matter perturbations, represented through the fractional matter perturbation Eq.~\eqref{eq:fractional_matter_perturbation}, illustrates the growth of large scale structures in the Universe through the M\'esz\'aros equation of the form given in Eq.~\eqref{eq:meszaros_form}. Model independent surveys of large scale structures~\cite{Abdalla:2022yfr,DiValentino:2020vvd} show an increasing tension with the $\Lambda$CDM predictions in the concordance model~\cite{Copeland:2006wr}, resulting in the consideration of modified cosmologies. Such models will result in different evolution scenarios of the large scale structure growth profile during the matter dominated era, where matter perturbations are related to the background equations yielding the expansion of the Universe. Here we considered a scalar-tensor model by including a potential and kinetic term in addition to the Einstein-Hilbert action, as shown in Eq.~\eqref{eq:action}.

The growth of large scale structures is correlated to the scalar cosmological perturbations of the gravitational sector. Gravitational scalar perturbations are given by the metric tensor in Newtonian gauge in Eq.~\eqref{eq:metric_perturbation} along with perturbation of the scalar field in Eq.~\eqref{eq:field_perturbation}, while matter perturbations are represented by Eq.~\eqref{eq:energy_momentum_perturbation} through the energy-momentum tensor. Variations with respect to the metric $g_{\mu\nu}$ and scalar field $\phi$ yields the to the background Friedmann equations~(\ref{eq:field_equations_background_W00}-\ref{eq:field_equations_background_Wii}) and Klein-Gordon equation~\eqref{eq:field_equations_background_Wphi}, respectively. Additionally, the first order perturbation of the equations of motion~(\ref{eq:field_equations_perturbations_W00}-\ref{eq:field_equations_perturbations_Wphi}) along with the conservation of energy-momentum tensor in Eqs~(\ref{eq:continuity_perturbation}-\ref{eq:velocity_perturbation}) provides a system of equations to formulate the evolution of scalars.

When observing the growth of large scale structure, it suffices to consider the matter epoch where CDM dominates. For this reason, the anisotropic stress listed in Eq.~\eqref{eq:energy_momentum_perturbation} is eliminated as quadrupole moments are prevalent in radiation dominated era. Moreover, the assumption in Eq.~\eqref{eq:CDM_conditions} where both background and perturbed pressure is set to zero. In $f(R)$ gravity it has been shown that performing the growth analysis using a generalized M\'esz\'aros equation dependent on the spatial wave mode $k$~\cite{delaCruz-Dombriz:2008ium} yields a distinct result for $G_{\text{eff}}$ in the quasi-static limit from that obtained where time derivatives of potentials are ignored~\cite{Tsujikawa:2007gd,DeFelice:2010aj-f(R)_theories}. Hence, the immediate assumption of the subhorizon limit $k \gg aH$ to construct the M\'esz\'aros equation could oversimplify the result. On the other hand, in the case of torsional $f(T)$ gravity, it has been shown that such distinction does not always occur~\cite{Zheng:2010am,Capozziello:2023giq}. For this reason, we considered $k$-dependency throughout the analysis, along with the comparison of the solution obtained through the subhorizon limit assumption.

In this paper, we consider a variety of potential models while setting $P(\phi) = 1$. The M\'esz\'aros equation~\eqref{eq:meszaros_in_a} is shown to be independent of the potential. Regardless, the different potentials dictate the behaviour of the background equations. In Sec.~\ref{sec:models} we consider a quadratic, quartic and exponential potential. It should be noted that the current growth index Eq.~\eqref{eq:growth_index_current} is independent of the potential and the behaviour of background equations. This is due to the fact that normalized Hubble parameter $h(1) = 1$, and the value of $\phi'(1) = 0$ is determined by setting the effective equation of state~\eqref{eq:effective_EoS} to approach $\omega_{\text{eff}} = -1$. Although different boundary conditions can be considered for $\phi'(1)$, the results do not satisfy a viable background cosmology within this framework. Fig.~\ref{fig:GrowthIndex_Case1to3} illustrates the solution for a linear dynamical growth index $\gamma$ in order to obtain the $\gamma_{\Lambda\text{CDM}}$ solution~\cite{Planck:2018vyg}, as summarized by Table~\ref{tab:Case1to3_gamma}. It was shown that the solution is $k$-dependent and only approaches the subhorizon solution for $k > 500 H_{0}$.

For the quadratic potential in Sec.~\ref{sec:Case1}, a variety of $\phi_0$ values are considered, for which the effective equation of state is illustrated in Fig.~\ref{fig:omega_effective_Case1} where viable models are for $10 < \phi_0 \leq 100$. The deviation of the growth factor from $\Lambda$CDM solution is illustrated in Fig.~\ref{fig:Growth_Factor_Case1} for which the viable model in Fig.~\ref{fig:GrowthFactor_Case1_100} shows a faster growth. For very large values of $k$, we can see a solution approaching that of the subhorizon limit. Similarly, the fourth power potential provides a larger range of viable values within the range $1 < \phi_0 \leq 100$ as seen in Fig.~\ref{fig:omega_effective_Case2}. As $\phi_0$ values increase, growth factor solutions are seen to grow faster than the $\Lambda$CDM result. In contrast, Fig.~\ref{fig:GrowthFactor_Case2_1} shows a slower growth than $\Lambda$CDM limit. This particular subcase can be disregarded due to the oscillatory behaviour of $\omega_{\text{eff}}$. 

The last case considered is for the exponential model in Sec.~\ref{sec:Case3}. Initially, $\lambda$ is set to be unitary to observe the behaviour for a variety of $\phi_0$ values. Fig.~\ref{fig:omega_effective_Case3} indicates that there is insignificant difference in the choice of $\phi_0$ and also reflected in the quasi-identical behaviour of Figs~\ref{fig:Growth_Factor_Case3}. Fig.~\ref{fig:omega_effective_Case4} depicts that for $\lambda > 0$ gives a dynamical model, whereas an increasing $\lambda$ value yields a model approaching a stiff behaviour. With the exception of $k = 100 H_0$ in Fig.~\ref{fig:GrowthFactor_Case4_100} where growth is slower, the rest of the models in Fig.~\ref{fig:Growth_Factor_Case4} illustrate a faster growth as $\lambda$ increases. It should be noted that the case $\lambda = 0$ yields a growth factor solution mimicking the behaviour given by $\Lambda$CDM.

Across all cases, it is evident that growth factor and growth index solutions are $k$ dependent. The expectation of solutions coinciding with the subhorizon solution can only be achieved for very large values of $k$. However, it should be noted that even $k = 100 H_0$ should attain the subhorizon limit, as shown in Ref.~\cite{Capozziello:2023giq}. It would be interesting to consider a broader class of models and carry out a comparison with observational data. The M\'{e}sz\'{a}ros equation may provide a key piece of information on which to build viable alternatives to the standard model of cosmology. Moreover, scalar-tensor cosmological models provide a vast range of model-space possibilities on which to build cosmological models. Finally, such a framework provides the building blocks to analyse growth in the radiation epoch in the super-horizon limit along with the consideration of stress anisotropies.

\acknowledgments{The authors would also like to acknowledge funding from ``The Malta Council for Science and Technology'' in project IPAS-2020-007. This paper is based upon work from COST Action CA21136 {\it Addressing observational tensions in cosmology with systematics and fundamental physics} (CosmoVerse) supported by COST (European Cooperation in Science and Technology).  MC  acknowledges funding by the Tertiary Education Scholarship Scheme (TESS, Malta). JLS would also like to acknowledge funding from ``The Malta Council for Science and Technology'' as part of the REP-2023-019 (CosmoLearn) Project.}

\bibliographystyle{utphys}
\bibliography{references}

\appendix

\section{Coefficients of Field Equations} \label{sec:AppendixA}

The coefficients of the gravitational perturbations and of the fractional matter perturbation in field equations and their time derivatives are given in their generalised forms in Eqs~(\ref{eq:A1}-\ref{eq:A8}). The system is set up as matrix notation expressed in Eq.~\eqref{eq:matrix}. The linearly independent set of equations allow for the invertibility of the matrix to obtain
\begin{align}
    \mathcal{X} = \mathcal{B}^{-1}\mathcal{C}\,,
\end{align}
where $\mathcal{X}^{\top} = (\delta\phi, \dot{\delta\phi}, \ddot{\delta\phi}, \varphi, \dot{\varphi}, \psi, \dot{\psi}, \ddot{\psi})$, to simultaneously solve each variable and its derivative in terms of $(\delta_m, \dot{\delta}_{m}, \ddot{\delta}_{m})$. The following functions found in Eqs~(\ref{eq:A1}-\ref{eq:A8}) correspond to the metric and scalar field equations in Eqs~(\ref{eq:field_equations_perturbations_W00}-\ref{eq:field_equations_perturbations_Wphi}) by eliminating the variable $v$ and $\delta\rho$ through the substitution of Eq.~\eqref{eq:v_equation} and Eq.~\eqref{eq:fractional_matter_perturbation}, respectively:
\begin{align}
    \mathcal{A}_{1} &= \kappa^2 \rho \left( \delta_{m} + \tfrac{3 H}{\tfrac{k^2}{a^2} - 3 \dot{H}} (\dot{\delta}_{m} - 3 H\varphi + 3\dot{\psi})\right) + \tfrac{1}{4}(2V_{\phi} + P_{\phi}\dot{\phi}^2) \delta\phi - \tfrac{1}{2} P \dot{\phi} (\dot{\phi} \varphi- \dot{\delta\phi}) \nonumber \\ & \quad + 2 \tfrac{k^2}{a^2} \psi  + 6 H (H \varphi - \dot{\psi}) \,, \\
    \mathcal{A}_{2} &= - P \dot{\phi} \delta\phi + 4 (H\varphi + \dot{\psi}) + \tfrac{2 \kappa^2 \rho}{\tfrac{k^2}{a^2} - 3 \dot{H}} (\dot{\delta}_{m} - 3 H \varphi - 3 \dot{\psi})\,, \\
    \mathcal{A}_{3} &= 6 \big[( V_{\phi} - \tfrac{1}{2} P_{\phi} \dot{\phi}^2)\delta\phi - P \dot{\phi} \dot{\delta\phi} + (12 H^2 + 8 \dot{H} + P \dot{\phi}^2) \varphi + \tfrac{4}{3} \tfrac{k^2}{a^2} (-\varphi + \psi) \nonumber \\ & \quad + 4 H (\dot{\varphi} + 3 \dot{\psi}) + 4 \ddot{\psi} \big] \,, \\
    \mathcal{A}_{4} &= \varphi - \psi \,, \\
    \mathcal{A}_{5} &= \tfrac{1}{2}\left( 2\tfrac{k^2}{a^2} - \tfrac{P_{\phi}}{P} (2 V_{\phi} + P_{\phi}\dot{\phi}^2) + 2 V_{\phi\phi} + P_{\phi\phi}\dot{\phi}^2 \right) \delta\phi + (3 H P + P_{\phi} \dot{\phi}) \dot{\delta\phi} + P \ddot{\delta\phi} + 2 V_{\phi} \varphi \nonumber \\ & \quad - P \dot{\phi} (\dot{\varphi} + 3 \dot{\psi})\,, \\
    \mathcal{A}_{6} &= \dot{\mathcal{A}}_{1}\,, \\
    \mathcal{A}_{7} &= \dot{\mathcal{A}}_{2}\,, \\
    \mathcal{A}_{8} &= \dot{\mathcal{A}}_{4}\,.
\end{align}
All coefficients in matrices $\mathcal{B}$ and $\mathcal{C}$ can be constructed through these functions.

\end{document}